\documentclass[epj]{svjour}
\usepackage{graphicx,epsf,amsmath,amssymb,wasysym,verbatim,color}

\begin{document}

\title{Identifying an influential spreader from a single seed in complex networks via a message-passing approach}
\author{Byungjoon Min\inst{1}}
\institute{
	IFISC, Instituto de F\'isica Interdisciplinar y Sistemas Complejos (CSIC-UIB),
	Campus Universitat Illes Balears, E-07122 Palma de Mallorca, Spain
  \email{byungjoon@ifisc.uib-csic.es}
}
\date{Recevied: date / Revised version: date}
\abstract{
Identifying the most influential spreaders is one of outstanding problems
in physics of complex systems. So far, many approaches have attempted to rank
the influence of nodes but there is still the lack of accuracy to single 
out influential spreaders. Here, we directly tackle the problem of finding 
important spreaders by solving analytically the expected size of epidemic 
outbreaks when spreading originates from a single seed. We derive and 
validate a theory for calculating the size of epidemic outbreaks with a single 
seed using a message-passing approach. In addition, we find that the probability 
to occur epidemic outbreaks is highly dependent on the location of the seed 
but the size of epidemic outbreaks once it occurs is insensitive to the seed. 
We also show that our approach can be successfully adapted into weighted networks. 
\PACS{
{05.70.Ln}{Nonequilibrium and irreversible thermodynamics}
\and{89.65.-s}{Social and economic systems}
\and{89.75.Hc}{Networks and genealogical trees}
}
}

\authorrunning{B. Min}
\titlerunning{Identifying an influential spreader from a single seed via a message-passing approach} 
\maketitle

\section{Introduction}
The topological location of the origin of spreading dynamics plays an 
important role in a final configuration of spreading processes
\cite{rmp,kitsak,castellano,moreno,klemm,control,target,pei}. For example, an initial 
patient from which an epidemic starts to spread influences critically the 
total number of infected patients~\cite{kitsak,radicchi,lu}. In addition, 
influential spreaders are required to be targeted for immunization with 
a high priority to halt epidemic outbreaks or rumor spreading 
\cite{cohen,holme,chen,masuda,altarelli}. Searching for the most influential nodes 
in complex networks has attracted much attention from many disciplines 
such as physics, complex network science, sociology, and computer science 
due to its practical application in real-world spreading processes including 
emerging epidemics and information diffusion~
\cite{kitsak,castellano,moreno,klemm,pei,radicchi,lu,bauer,pei2,bmin,morone,morone2,morone3,liu,malliaros}.

Several centralities in terms of network topology from degree~\cite{albert}
that is the number of neighbors to $k$-core~\cite{kcore}, betweenness 
centrality~\cite{bc}, and PageRank~\cite{pagerank} have been tested 
for identifying influential spreaders. Even though many methods so far have
been proposed to single out influential spreaders, however there are still
limitations in their accuracy and applicability because most of them were 
based on intuition rather than on mathematical background. 
In addition, while the ranking of nodes' influence would rely on the details of spreading
processes such as the probability of transmission, most previous methods do 
not take into account the processes of spreading dynamics 
explicitly~\cite{kitsak,radicchi,lu,bmin,morone}.

In order to overcome these limitations, we directly derive a theory for
finding influential spreaders based on a message-passing approach. 
We compute the expected size of epidemic outbreaks 
on locally tree-like networks, when an epidemic starts from a single node.
We validate our theory with extensive numerical simulations on synthetic 
and empirical networks with various transmission probabilities. We confirm 
that our theory can predict accurately the influence of 
spreaders in complex networks. We also find that the location of an initial 
spreader affects the probability of epidemic outbreaks but not the average size 
of epidemic outbreaks once it occurs. In addition, we show that our approach 
based on message-passing equations can be applied to weighted networks.

\section{Theory}

We consider the susceptible-infected-recovered (SIR) model as a typical epidemic 
model \cite{kermack}. The SIR model is regarded 
as one of the most simple yet successful models describing irreversible spreading 
processes. The SIR model consists of three states: susceptible (S), infected (I), 
and recovered or removed (R). Each infected node is infectious to 
spread disease to its neighbors on a network with the infection rate $\beta$. 
Independently, each infected node becomes recovered after the recovery 
time $\tau$. Once recovered, it is not infectious anymore, leading 
to an irreversible process. 
Thus, an infected node is able to spread a disease 
to its neighbors during the recovery time from the moment of infection. 
For the sake of simplicity, we assume that the 
recovery time is sharply distributed, and so its probability distribution 
$P(\tau)$ is well described by the delta function.

We implement the SIR model on complex networks with a single seed. To be specific, 
all nodes are initially susceptible except a single infected node $i$ which 
corresponds to the first patient. An infected node transmits 
disease to its neighbors with the infection probability $\beta$ and 
autonomously recovers with the recovery time $\tau$. Without loss of generality,
we set $\langle \tau \rangle = 1$ in our study. These processes proceed until 
there are no more infected nodes in a system. In the stationary state ($t \rightarrow \infty$),
we measure the fraction of recovered nodes $\rho_i$ when epidemic starts 
with a single seed $i$, called the prevalence of epidemic outbreaks.
The higher prevalence $\rho_i$ is the higher influence of node $i$ is,
because higher $\rho_i$ implies that an epidemic initiated by node $i$ 
brings out larger epidemics in average.

Once we obtain the prevalence of each node $\rho_i$, we can directly identify 
influential spreaders based on sorting of $\rho_i$. However, it is a 
time-consuming process to obtain $\rho_i$ for every node in a system 
by numerical simulations because many realizations for every different seed 
are needed. Thus, we derive a theory for estimating the prevalence for a seed
via a message-passing approach. A message-passing approach has also been applied 
for percolation \cite{karrer}, inferring the origin of spreading \cite{lokhov2},
optimal immunization \cite{altarelli},
optimal deployment of resource \cite{lokhov3}, and 
optimal percolation \cite{morone,braunstein}.
We use mapping between the epidemic model with 
the static bond percolation, which is well known for a long 
time~\cite{grassberger,cardy,newman}. Then, we reinterpret the message-passing 
approach for bond percolation as a theory for identifying superspreaders in the 
epidemic model \cite{radicchi}. In this section, we present the message-passing equations and 
its interpretation for the SIR model.

We first consider a classical bond percolation problem on complex networks with a size 
$N$ and link occupation probability $T_B$. In a percolation process, there
can be one giant component $\mathcal{G}$ that is a connected cluster that covers 
a non-vanishing fraction of a network in the limit $N \rightarrow \infty$ 
and multiple small components. The giant component appears only if $T_B$ is 
sufficiently high, i.e., $T_B$ is larger than the percolation threshold. If 
$T_B$ is less than the threshold, there exist only small components. At 
the percolation threshold, the giant component appears, showing a typical 
second-order phase transition between non-percolating and percolating phases.

For the mapping between the bond percolation with SIR model, let us imagine 
a final configuration of the SIR model in the steady state. In the limit 
$t \rightarrow \infty$, all nodes in a network are either susceptible 
or recovered. Note that all infected nodes become eventually recovered. 
We then define links on which infection occurs as infection links
whose probability is given by transmissibility $T$.
The transmissibility is then the probability that an infected node infects 
its neighbors before it recovers, and therefore 
$T = 1 - e^{-\beta \tau}$~\cite{newman}.
The relation between the SIR model and bond 
percolation is established by correspondence between the transmissibility $T$ 
in the SIR model and the probability of link occupation in bond percolation 
$T_B$. Then, the size of epidemic outbreaks in the steady state corresponds to 
the size of connected component by infection links which is fully determined by 
the structural property of a final configuration. In the steady state, the epidemic 
size at a given $T$ corresponds to the size of $\mathcal{G}$ 
with the same occupation probability $T_B$ in the bond percolation jargon. 
Thus, a continuous transition between disease-free phase and global epidemic phase 
at the epidemic threshold is observed, similar with the bond percolation.
Based on this mapping, we derive a theory for finding influential nodes in
the SIR model, using a message-passing approach developed in percolation theory.

In order to derive the theory for the SIR model, first define $H_{ij}$ 
as the probability that node $j$ by following a link from node $i$ does not 
bring out an epidemic outbreak with the transmissibility $T$. Assuming 
locally tree-like structures, if neither the link between $i$ and $j$ 
is infection link nor all the neighbors of node $j$ excluding node $i$ 
occur an epidemic outbreak, node $j$ does not bring out an epidemic outbreak. 
Thus, the probability $H_{ij}$ can be obtained by following coupled self-consistency 
equations \cite{radicchi,karrer,karrer2,lokhov,radicchi2},
\begin{equation}
H_{ij} = 1 - T + T \prod_{k \in \partial j \setminus i} H_{jk}.
\label{sir}
\end{equation}
where $k \in \partial j \setminus i$ represents a set of neighbors of node $j$ 
excluding node $i$. 
Computing the above self-consistency equations iteratively, $H_{ij}$ converges 
towards a fixed point. 
We here concentrate only on locally tree-like structure, but our framework can 
also be extended for networks with triangles beyond locally tree-like 
structures \cite{radicchi2}.

\begin{figure*}
\begin{center}
\includegraphics[width=0.85\linewidth]{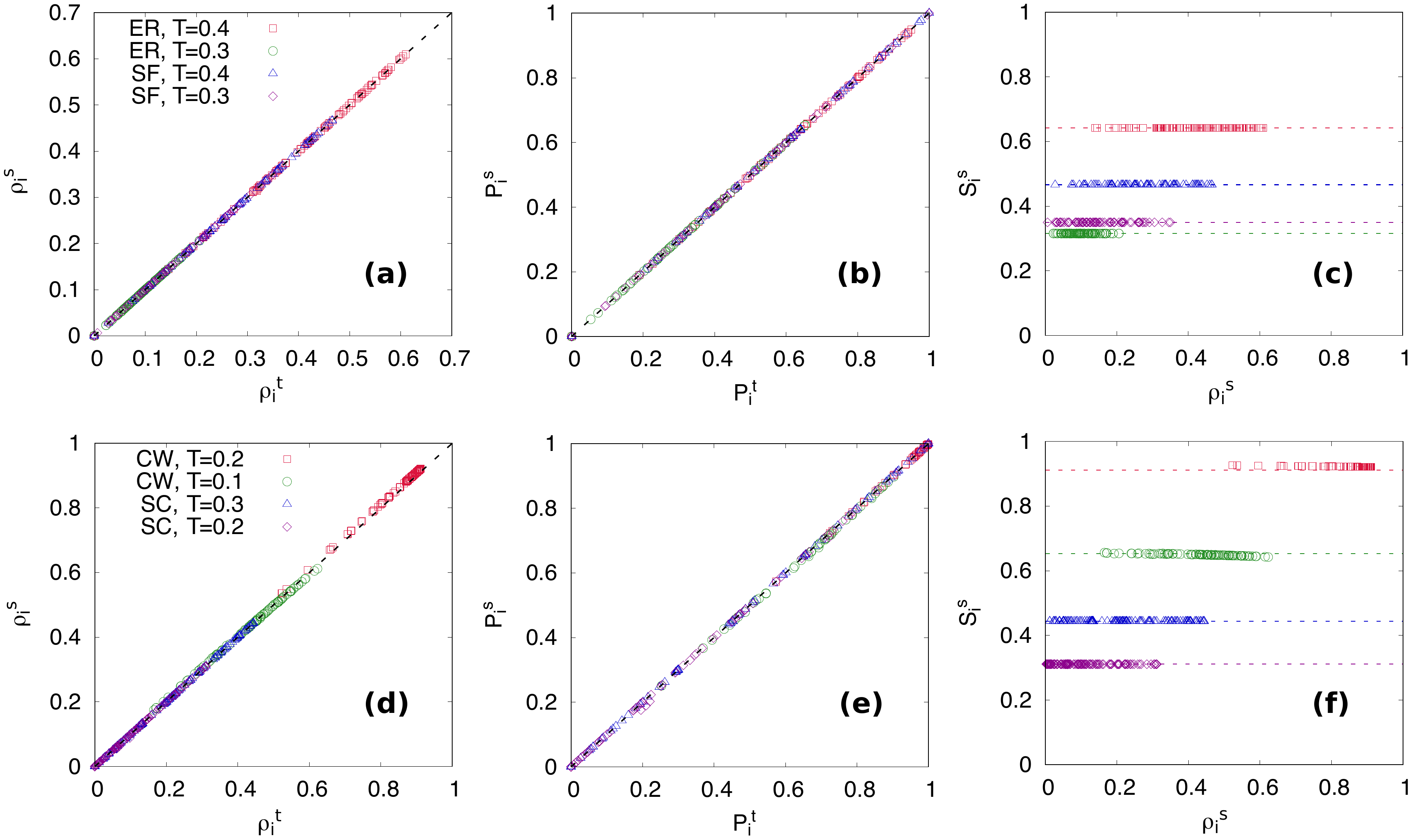}
\caption{
Scatter plot of (a) prevalence $\rho_i$, (b) probability $P_i$,
and (c) size $S_i$ of epidemics initiated from node $i$
obtained by theory and numerical simulations 
for ER and SF networks with $N=10^5$ with different transmissibility.
The same plot for the empirical networks, CW and SC, are shown in (d-f). Dashed lines 
in (c) and (f) represent average epidemic size obtained by theory 
as $S = \frac{1}{N} \sum_i^N P_i$.
The theory is in perfect agreement with the numerical simulations 
for both random and empirical networks.
}
\end{center}
\end{figure*}

When $H_{ij}$ is obtained, we can calculate the probability 
$P_i$ that a seed $i$ triggers an epidemic outbreak in terms of $H_{ij}$ as
\begin{equation}
P_i = 1 - \prod_{j \in \partial i} H_{ij},
\label{pi}
\end{equation}
where $j \in \partial i$ indicates a set of neighbors of node $i$. When $T$ is 
less than the epidemic threshold, $P_i$ is zero since there is no global epidemic.
In the bond percolation jargon, $P_i$ is the probability that a randomly chosen 
node $i$ belongs to the giant component $\mathcal{G}$.
Similarly, we can obtain the size of epidemic when it occurs from a seed $i$ by
\begin{equation}
S_i = \frac{1}{N} \left(1+\sum_{\substack{j=1 \\ j\ne i}}^N P_j \right).
\label{sirsize}
\end{equation}
Since a node $i$ has to be included in the epidemic outbreak as a seed, we 
differently treat node $i$ in the summation. 
After we obtain the probability and size of epidemic outbreaks, we can simply 
calculate the average prevalence when the epidemic is initiated by a seed $i$ as
\begin{equation}
\rho_i = P_i S_i. 
\label{rho}
\end{equation}
In summary, (i) we obtain $H_{ij}$ by calculating iteratively Eq.~1
with a given $T$, (ii) compute $P_i$ by Eq.~2, $S_i$ by Eq.~3, 
and $\rho_i$ by Eq.~4 for different seed selection, and 
(iii) identify superspreaders based on sorting the prevalence of each node $\rho_i$.

Our theory provides not only the ranking of influential spreaders but also 
intriguing perspectives on finding superspreaders. First, the size of epidemic 
outbreaks is insensitive to the location of an initial seed once the spread 
reaches global epidemics. Thus the difference in the influence for different 
seeds is mainly caused by the epidemic probability of each seed. 
Second, the ranking of influential spreaders is not fully determined 
by only the topological location of seed but can vary depending on 
the parameters of spreading processes and details of spreading models~\cite{radicchi3}.
Therefore, it would be misleading if one attempts to find a universal 
ranking of influential spreaders solely relying on network structures
ignoring dynamical properties of spreading processes as reported in \cite{radicchi3}.

\section{Results}

We first test our theory with numerical simulations on Erd\"os-R\'enyi 
(ER) and scale-free (SF) networks with $N=10^5$ which respectively represent 
random graphs with a homogeneous and heterogeneous degree distribution. 
For building ER
networks, we randomly choose a pair of nodes and connect them unless
they are already connected. We repeat this step until the mean degree
$\langle k \rangle$ reaches the desired value. In our study, we set
$\langle k \rangle =4$.

For building SF networks, we use 
static scale-free network model \cite{goh}. In the model, each node 
$i$ has its inherent weight $\omega_i$ given by
$\omega_i = i^{-\mu} / \sum_{j=1}^N j^{-\mu}$, where $\mu$ is
a constant, $0<\mu<1$, which determines the degree exponent.
We choose a pair of nodes, say $i$ and $j$ independently 
following the probability $w_i$ and $w_j$ respectively,
and connect them unless they are already connected. 
We repeat this step until the mean degree $\langle k \rangle$ reaches the 
desired value which is $\langle k \rangle =2$ in our study. 
The degree distribution 
of the resulting network is asymptotically scale-free with the decaying 
tail $k^{-\gamma}$ with the degree exponent $\gamma=(\mu+1)/\mu$,
i.e., $\gamma=2.5$ in our study.

On the resulting network, we perform the SIR process with every single 
seed $i$. The prevalence with seed $i$ is obtained,
averaged over $10^4$ independent runs. The prevalence obtained by the theory 
$\rho_i^{t}$ and numerical simulation $\rho_i^{s}$ are shown together
in Fig.~1(a). 
Our theory exhibits perfect agreement with the numerical results. 
Pearson correlation coefficient between $\rho_i$ for the theory 
and simulation is larger than $0.99$ 
for all tested infection rate and network structures.
Note that non-backtracking centrality corresponds to the limit
$H_{ij} \rightarrow 1$ where $T$ is at an epidemic threshold. Thus, 
non-backtracking centrality can predict the influence of spreaders 
reliably at the epidemic threshold as a special case of our theory 
near the epidemic threshold \cite{radicchi,martin}.

\begin{figure*}
\begin{center}
\includegraphics[width=0.8\linewidth]{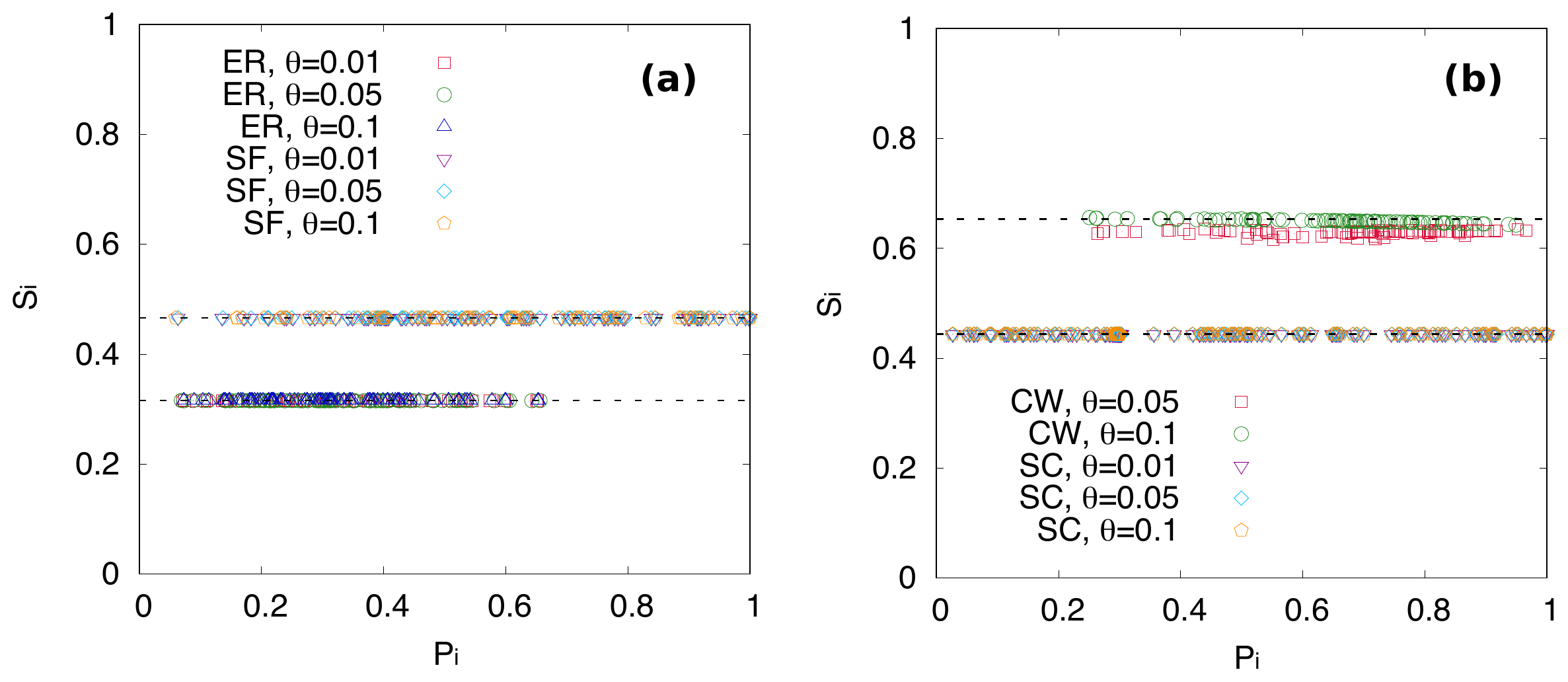}
\caption{
(a) Scatter plot of size $S_i$ of epidemics initiated from node $i$
with respect to probability of epidemics $P_i$
obtained by the numerical simulations
for ER ($T=0.3$) and SF ($T=0.4$) networks with $N=10^5$
and different threshold $\theta=0.01,0.5,0.1$.
(b) The same plot for the empirical networks, CW ($T=0.1$) and SC ($T=0.3$), 
with $\theta=0.01,0.5,0.1$. 
For the CW network, the result with $\theta=0.01$ is discarded
since the network size is less than 100. 
Dashed lines represent average epidemic size obtained by theory. 
}
\end{center}
\end{figure*}

We also validate our theory for the SIR model on top of empirical contact 
networks. In order to reconstruct real-world networks, we use contacts in a 
workplace network~(CW) from face-to-face contact patterns between individuals 
in an office building in France \cite{genois} and a sexual contact network (SC) 
containing the information of sexual activity gathered from a web community 
of internet-mediated prostitution in Brazil \cite{rocha}. Both networks 
contain the moment of contacts between individuals. Note that both networks 
are far different from random networks because the CW has a strong
modular structure based on the organization of the offices in departments
and the SC is a completely bipartite network. Therefore, these networks 
are suitable examples to evaluate how well the theory works beyond random 
networks. As shown in Fig.~1(d), the theory predicts the final fraction of 
epidemic outbreaks accurately for the empirical networks. Correlation
coefficient between the theory and numerical simulation 
reaches more than $0.99$, indicating perfect agreement between them.

Beyond the agreement in the prevalence between the theory and numerical results,
we decompose $\rho_i$ into the probability $P_i$ and size $S_i$ of epidemics.
In numerical simulation, we define global epidemics when more than 10~\% 
of nodes in a network ultimately are infected.
We compute numerically the probability $P_i$ of epidemic outbreaks 
as the frequency of global epidemics out of all trials
and the size $S_i$ of outbreaks when global epidemic occurs. 
We compare the probability and size obtained by the theory and numerical 
results for ER, SF, CW, and SC networks in Fig.~1. 
We find that our theory reliably predicts the probability and size of 
epidemic outbreaks. We also find that while the probability $P_i$
shows different values for different initial seeds, the size $S_i$ is almost 
constant and insensitive to seed location. Thus, the epidemic size once global 
epidemic occurs $S_i$ is independent to $\rho_i$.
This result implies that the location of seed affects the probability 
to bring out a global epidemic but not the size of an epidemic once it occurs.

We also confirm that the numerical results are highly robust for the different 
definition of the global epidemic. Specifically, we check the size of 
epidemics $S_i$ when the fraction of infected nodes exceeds a threshold value $\theta$,
1, 5, and 10~\% out of all nodes (Fig.~2). We find that the different 
threshold values $\theta$ do not produce notable difference at all.
The irrelevance of the seed location to $S_i$ can be understood 
if we recall the mapping between the bond percolation and the SIR model.
The epidemic size $S_i$ from a single seed can be regarded as the average size 
of giant component including node $i$ in the perspective on the bond percolation.
Therefore, $S_i$ is rather insensitive to the topological location of a seed $i$.
In contrast, $P_i$ strongly relies on the location of a seed 
since it can be interpreted as the probability that a randomly chosen 
node belongs to the giant component in the bond percolation.

Next, we generalize our theory for weighted networks, assigned different transmissibility
for each link. Despite ubiquity of heterogeneity in link weights, most centralities 
of complex networks have
aimed to single out influential spreaders on unweighted networks except 
a few studies~\cite{garas,gao}.
We here modify the theory for unweighted networks by introducing 
a different transmissibility $T_{ij}$ for each link. Substituting $T_{ij}$ into $T$ in Eq.~1,
we can obtain the probability $H_{ij}$ that node $j$ arrived through a link 
from node $i$ does not bring out epidemic outbreaks on weighted networks by
\begin{equation}
H_{ij} = 1- T_{ij} + T_{ij}	\prod_{k \in j \setminus i} H_{jk}.
\label{eq1}
\end{equation}
After getting $H_{ij}$, the probability that a seed node $i$ produces  
epidemics and the average size of epidemics when a global epidemic occurs 
can be computed by using the same equations for unweighted networks.

We test the theory for weighted networks with CW and SC networks. Both datasets
contain the number of contacts between two individuals, $\nu_{ij}$. 
We define the transmissibility between individuals as $T_{ij}=1-e^{-\beta \nu_{ij} \tau}$.
The transmissibility increases with increasing $\nu_{ij}$,
and $T_{ij}$ becomes unity in the limit $\nu_{ij} \rightarrow \infty$. 
When all $\nu_{ij}=1$, it is reduced into an unweighted network.
Taking into account the weights of each link, we calculate the prevalence
of epidemic outbreaks for each seed on CW and SC networks. As shown in
Fig.~3, we find that our theory for weighted networks is in well agreement 
with the numerical simulations for the prevalence of outbreaks.
Thus, we confirm that our method based on message-passing equations 
can also predict reliably the influential spreaders in weighted networks.

While our theory is reliable for the identification of an influential
spreader on locally tree-like networks, it has some limitations.
First, the theory might break down for networks with many short loops such as 
spatial networks~\cite{spatial} since they violate the assumption of locally 
tree-like structures. Second, our method does not guarantee its validity for
finding a set of multiple influential seeds that can spread disease 
or information to the largest part of a network. Note that 
the problem of identifying multiple influential spreaders is far different 
and difficult than that of a single spreader because infected nodes
by different seeds can be largely overlapped \cite{pei,kempe}.

\begin{figure}
\begin{center}
\includegraphics[width=0.8\linewidth]{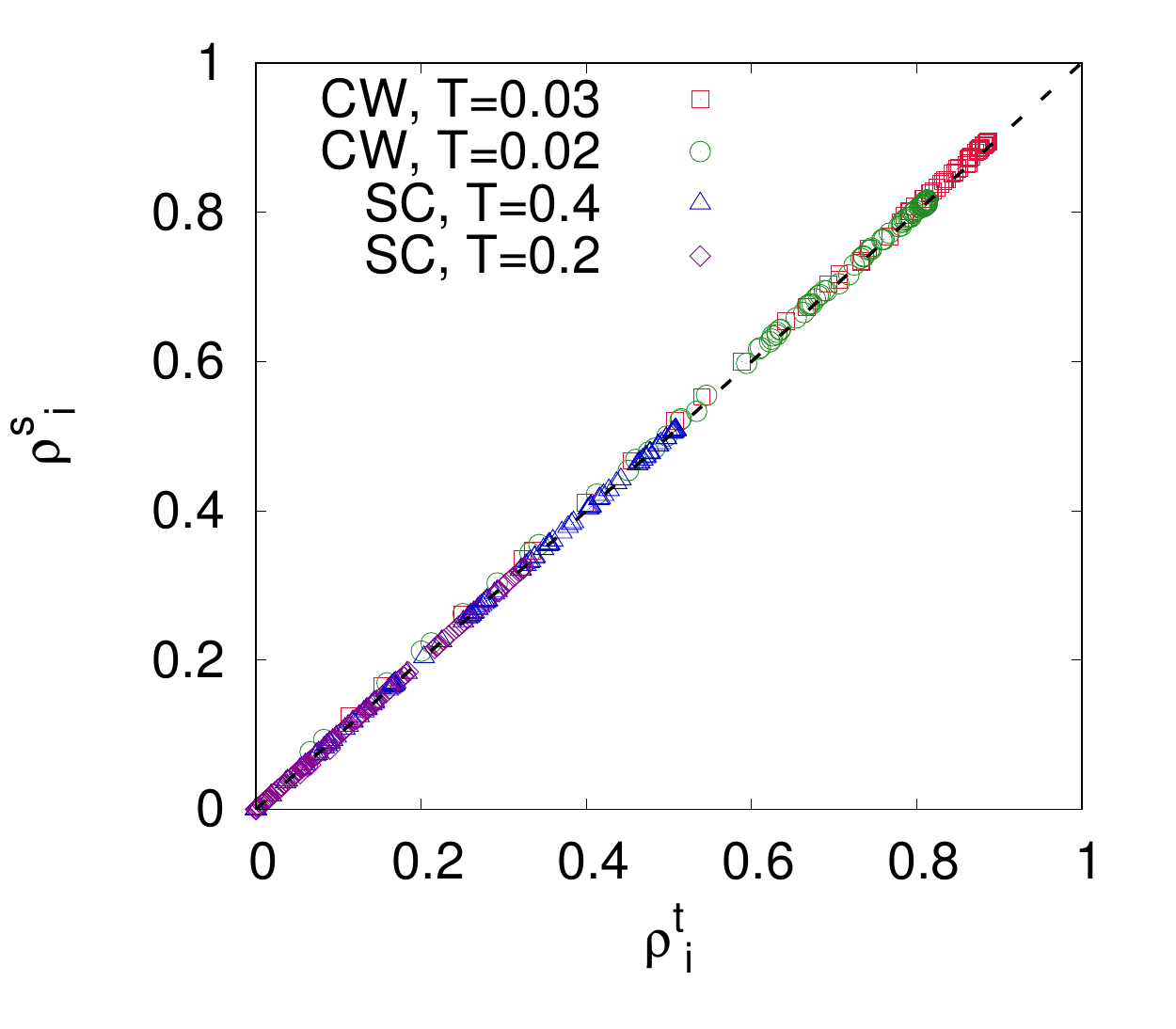}
\caption{
Scatter plot of the prevalence for theory $\rho_i^{t}$ 
and numerical simulations $\rho_i^{s}$ on top of weighted networks obtained 
from CW and SC networks. The theory is in good agreement with the numerical simulations.
}
\end{center}
\end{figure}

\section{Discussion}

In this study, we propose an accurate method for identifying the most influential 
spreaders for both unweighted and weighted networks via message-passing equations. 
Our theory is based on the exact mapping between the SIR model and bond percolation,
and exact for sparse networks with a locally tree-like structure. We show that
the theory is in perfect agreement with numerical calculations, and can outperform
previous approaches in terms of accuracy. Furthermore, we find that 
the location of seed affects the probability of epidemic outbreaks but not the size
of outbreaks, which is not well reported in previous study. 
Our study can shed light on the identification of the most important 
spreaders with a single seed in complex networks theoretically and practically, for 
instance for viral marketing, efficient immunization strategy, and identifying the 
most influential agents in society.

\begin{acknowledgement}
We thank M. San Miguel and R. Gallotti for useful discussions. 
We also thank A. Y. Lokhov for suggesting references related.
This work was supported by the Spanish  Ministry  MINEiCO  and  FEDER (EU) under the project 
ESOTECOS (FIS2015-63628-C2-2-R).
\end{acknowledgement}

\section*{Author contribution statement}
B. M. conceived the study, performed the research, analyzed data, 
and wrote the paper.

\end{document}